\documentstyle{article}
\begin{document}
\title{Feynman integral in regularized non-relativistic
quantum electrodynamics}
\author{Z. Haba\\Institute of Theoretical Physics,\\ 
University of
Wroclaw, Wroclaw, Poland}
\date{}
\maketitle
\begin{abstract}
We express the unitary time evolution in  non-relativistic
regularized quantum electrodynamics at zero and positive temperature
by a Feynman integral defined in terms of a complex Brownian motion.
An average over the quantum electromagnetic field determines the form
of the quantum mechanics in an environment of a quantum black body
radiation. In this non-perturbative formulation of quantum
electrodynamics we prove the existence of the classical limit
$\hbar \rightarrow 0$.We estimate an error to some approximations
commonly applied in quantum radiation theory.
\end{abstract}
\vfill
\hfill {\Large \bf March 1996 ITP UWr 905/96}
\thispagestyle{empty}

\newpage
\section{Introduction}
Quantum electrodynamics (non-relativistic as well as relativistic)
has been formulated in terms of a path integral by Feynman himself
 \cite{fey1}. Starting from the Feynman integral the conventional
 perturbation series in powers of the fine structure constant
 $\alpha=e^{2}(\hbar c)^{-1}  $ is derived. It is a common attitude
 among the physicists to treat the Feynman integral just as a tool
 for a generation of the perturbation series. However, it is well
 known that the imaginary time (Euclidean) version of the Feynman
 integral has a mathematical meaning beyond the perturbation theory.
 In fact, the Euclidean two-dimensional quantum electrodynamics (QED)
 has been constructed by means of the path integral(\cite{cha}
 \cite{bry}). The path integral for Euclidean regularized QED at finite
  temperature
 in any dimension has been discussed in ref.\cite{roe}. Although, in
 principle, the Euclidean correlation functions
  can be continued analytically to the Minkowski space \cite{ost}, in
 practice, this is an inconvenient  approach to dynamics.

 Perturbative QED agrees perfectly well with experiments.
 Its non-perturbative effects have not been discovered yet.However,
 the photons form an inevitable environment for all events in nature
 at least in the form of the heat or (much weaker) cosmic microwave
 background. The permanent influence of radiation may have a
  profound effect on mesoscopic phenomena. A lot of theoretical
  studies has been done investigating the effect of an
  infinite environment on a finite microscopic system (see e.g.
  \cite{for} \cite{leg2}
  \cite{leg1}).
  However, the mathematical model of the heat bath could not be identified
   with any physical environment. The quantum electromagnetic field
   constitutes such a physical environment. There are suggestions
    \cite{cal}
   that Leggett$^{\prime}$s model approximates the effect of the quantum
   electromagnetic field. However, we think that this is quite a rough
   approximation completely out of control.The approximations
   commonly applied 
   \cite{pow}\cite{lui}\cite{spo}\cite{fro}\cite{feyver}\cite{naz}\cite{ued}
    neglect the spatial
   dependence of the electromagnetic field. The neglect of the spatial
   dependence is hard to justify in QED where the electromagnetic field
  has singular vacuum correlation functions.

In this paper we apply to QED our 
earlier mathematical formulation of the Feynman
 integral  \cite{hab2}-\cite{hab3} (see also \cite{Came} \cite{Doss}).
  Such a formulation makes it
  possible to discuss the dynamics
directly in the real time. Working with physical time is indispensable
in problems concerning finite time dynamics e.g. decoherence \cite{leg2}
\cite{zur} and radiation reaction. We consider the Dirac model
\cite{dir} of
a finite number of particles interacting by means of the Coulomb forces and
additionally with the quantum electromagnetic field. This is the first
approximation to the relativistic quantum field theory of interacting
quantum electromagnetic field (in the radiation gauge) and matter
fields. As pointed out already by Feynman \cite{fey1}a minor modification of
the path integral defines the complete relativistic model. For this
reason the non-relativistic theory has a similar perturbative expansion
in powers of the fine structure constant $\alpha$. At the moment we
are mainly interested in an application of the quantum electromagnetic
field as a model of an environment to the conventional non-relativistic
quantum mechanics. For this reason we include a discussion of mixed
states corresponding to a finite temperature. The main virtue of a
rigorous formulation of the Feynman integral is an opportunity to
estimate an error to formal calculations. In this paper we discuss
the radiation damping and quasiclassical expansion. In the perturbative
approach to QED we encounter a serious difficulty with the classical
limit (see \cite{bia}), because $\alpha$ tends to infinity when $\hbar
\rightarrow 0$. So, a non-perturbative formulation of this problem
is required.

The paper is organized as follows. In sec.2 we define the
model in terms of the Hamiltonian in a finite mode approximation
for the quantum electromagnetic field.In sec.3 we derive the
Feynman formula for the time evolution operator in this model.
The finite temperature QED is expressed by
the  functional integral in sec.4. In sec.5 we integrate over the
quantum electromagnetic field in order to describe the
quantum mechanical particles in equilibrium with a radiation background.
 We consider the limit of an infinite volume
and infinite number of modes in sec.6. We define the Feynman
kernel for the evolution operator in sec.7. We prove
a semiclassical asymptotics for the Feynman kernel in sec.8.
In sec.9 we discuss the approximation of a slowly varying
quantum electromagnetic field
(the dipole approximation) \cite{pow}\cite{lui}\cite{fro}\cite{spo}.
Finally, sec.10 contains a summary and perspectives for further
research.

We put more effort on the precise  description
of  quantum dynamics than on the details of the proofs.
This is so, because the method of the probabilistic representation
of the quantum dynamics has been described
in our earlier papers \cite{hab2}-\cite{hab3} together with detailed
proofs. Our approach in this paper starting with a finite mode
approximation is a repetition of methods of our earlier work.
\section{The Hamiltonian}
We consider the electromagnetic field in a finite volume $V$.We
expand the vector potential $A$ in eigenmodes $g(k,{\bf x})$
\begin{displaymath}
\omega(k)^{2}g(k,{\bf x})=c^{2}\triangle g(k,{\bf x})
\end{displaymath}
In this formula $\omega(k)$ denotes the eigenfrequencies and $k$
numerates the eigenmodes (so k is is a vector if V is a parallelepiped ;
this will be tacitly assumed further on, but we write some formulae in a way
suggesting  cavities of a general shape).
We assume that $g(k,x)$ are real (see e.g. refs.\cite{kro}\cite{lan}
for such a choice of eigenmodes) normalized as follows
\begin{displaymath}
\int_{V} g(k,{\bf x}) g(k^{\prime},{\bf x})dx=\delta(k,k^{\prime}) V
\end{displaymath}
The $\delta$-function is equal to 1 for the same $g$ and equal to zero
if the eigenmodes are different. The expansion in eigenmodes takes the
form
\begin{equation}
A_{j}(x)=\sum_{k,\nu}\sqrt{\frac{2\pi\hbar}{\omega(k)V}}
cf(k){\cal E}_{j}(k,\nu)(a(k,\nu)+a(k,\nu)^{+})g(k,x)
\end{equation}
In eq.(1)  ${\cal E}$ is the polarization transversal to $k$
\begin{displaymath}
\sum_{\nu=1,2} {\cal E}_{j}(k,\nu){\cal E}_{l}(k,\nu)\stackrel{\rm def}{=}
\delta^{tr}_{jl}=\delta_{jl}-k_{j}k_{l}|k|^{-2}
\end{displaymath}
$a(k,\nu)$ and $a(k,\nu)^{+} $ are the amplitudes fulfilling the
(Poisson brackets) commutation relations for creation and annihilation
operators.$f(k)$ is a formfactor regularizing the
electromagnetic field. Such a formfactor has been introduced already by
Pauli and Fierz
 \cite{fie} (we choose $f(k)$ real; with no regularization
$f(k)=1$).

The free Hamiltonian is
\begin{equation}
H_{R}=\sum_{k,\nu}\hbar\omega(k)a(k,\nu)^{+}a(k,\nu)=\sum_{k,\nu}
\frac{1}{2\mu(k)}p(k,\nu)^{2}+\frac{1}{2}\mu(k)\omega(k)^{2}y(k,\nu)^{2}
\end{equation}
where we defined
\begin{displaymath}
a(k,\nu)=\sqrt{\frac{\mu(k)\omega(k)}{2\hbar}} y(k,\nu)+
         \sqrt{\frac{1}{2\mu(k)\omega(k)\hbar}}ip(k,\nu)
\end{displaymath}
with
\begin{displaymath}
\mu(k)=\frac{4\pi f(k)^{2}}{\omega(k)^{2}V}
\end{displaymath}
In this way we have expressed the radiation Hamiltonian as a sum of
the Hamiltonians for harmonic oscillators. The Hamiltonian describing
an interaction of the r-th particle with an external potential $u_{r}$, the
interaction with photons (the coupling $\frac{e_{r}}{c}$, where $e_{r}$
 denotes
an electric charge and c the light velocity)
 and the Coulomb interaction among themselves
(these are the longitundinal degrees of freedom of the electromagnetic
field, hence the form of the regularized Coulomb interaction is a
consequence of eq.(1))reads
\begin{equation}
H_{P}=\sum_{r}\frac{1}{2m_{r}}({\bf p}_{r}+\frac{e_{r}}{c}{\bf A}({\bf
x}_{r}))^{2} +\sum_{r,s} e_{r}e_{s}\phi({\bf x}_{r},{\bf x}_{s}) +
\sum_{r}u_{r}({\bf x}_{r})
\end{equation}
where
\begin{displaymath}
\phi({\bf x}_{r},{\bf x}_{s})=4\pi\sum_{k}f(k)^{2}c^{2}\omega(k)^{-2}
g(k,{\bf x}_{r})g(k,{\bf x}_{s})
\end{displaymath}
When the number of modes is finite then the Hamiltonian
$H_{R}+H_{P}$ is defined as a self-adjoint operator in a
 tensor product of
a finite number of $L^{2}(dy)$ Hilbert spaces.
\section{Probabilistic representation of quantum dynamics}
If we restrict ourselves to a finite number of modes and quantize
the Hamiltonian $H=H_{R}+H_{P}$ canonically then we obtain a second
order (in ${\bf x}_{r}$ and $y(k,\nu)$) differential operator. A
probabilistic representation of the unitary group generated by $H$
has been discussed in our earlier papers \cite{hab2}.

Let us consider first a single harmonic oscillator.Denote its
Hamiltonian by $H_{k}$. The ground state $\chi_{k}$ of
$H_{k}$ is
\begin{equation}
\chi_{k}(y)=exp(-\frac{\mu_{k}\omega(k) y^{2}}{2\hbar})
\end{equation}
Let us define the Brownian motion as the Gaussian process with the
covariance
\begin{displaymath}
E[b_{t}b_{s}]=min(t,s)
\end{displaymath}
Let $q(t,y)$ be the solution of the stochastic equation
\begin{equation}
dq =-i\omega(k) qdt + \lambda \sigma_{k} db
\end{equation}
with the initial condition y. Here
\begin{displaymath}
\lambda =\frac{1}{\sqrt{2}}(1+i)
\end{displaymath}
and
\begin{displaymath}
\sigma_{k} = \sqrt{\frac{\hbar}{\mu(k)}}
\end{displaymath}

 Then,the Schr\"odinger time evolution of the state $\psi=\chi_{k}
\Phi$ can be expressed in the form
\begin{equation}
\psi_{t}(y)\equiv (exp(-iH_{k}\frac{t}{\hbar})\psi)(y)=\chi_{k}(y)
E[\Phi(q(t,y))]
\end{equation}
The free electromagnetic field is just a collection of harmonic oscillators.
Let $\chi $ be the ground state of $H_{R}$ i.e.$H_{R}\chi=0$.Let again
$\psi=\chi\Phi$.Then
\begin{equation}
\psi_{t}({\bf A})\equiv (exp(-iH_{R}\frac{t}{\hbar})\psi)({\bf A})=\chi({\bf
A})E[\Phi({\cal A}(t,{\bf A}))]
\end{equation}
where
\begin{displaymath}
{\bf A}=\sum_{k,\nu}\sqrt{4\pi\mu(k)}cf(k){\cal E}(k,\nu)y(k,\nu)
g(k,{\bf x})
\end{displaymath}
and
\begin{equation}
{\cal A}(t,{\bf A})=\sum_{k,\nu}\sqrt{4\pi\mu(k)}
cf(k){\cal E}(k,\nu)q(t;y,k,\nu)g(k,{\bf x})
\end{equation}
where
\begin{displaymath}
q(t;y,k,\nu)=exp(-i\omega(k)t)y(k,\nu)+\lambda \sigma_{k}
\int_{0}^{t}exp(-i\omega(k)(t-\tau))db(\tau;k,\nu)
\end{displaymath}
is the solution of eq.(5) with the initial condition $y(k,\nu)$
 (now each Brownian motion
has an index numerating the eigenmodes of the electromagnetic field;
the Brownian motions corresponding to different indices are
independent).

In order to derive a probabilistic representation of the Schr\"odinger
evolution for $ H_{R}+H_{P} $ we need first some additional cutoffs. So,
let us assume at the beginning that the number of modes is finite.Next,
we need  to bound the behaviour of ${\bf A}({\bf x})$ as well as
$\phi({\bf x}_{r},{\bf x}_{s})$ for complex ${\bf x}$.In ref.
\cite{hab2} we have bounded ${\bf A}$ and $\phi$ in the Feynman
integral multiplying
these functions by a Gaussian damping factor depending on the Brownian
 motion.  We introduce here a simpler
regularization particularly useful in QED. We regularize the modes
$g(k,{\bf x})$ as follows
\begin{equation}
g(k,{\bf x})\rightarrow g_{\delta}(k,{\bf x})=
exp(-\delta^{2} |{\bf x}|^{8})g(k,{\bf x})
\end{equation}
${\bf A}$ with the $g_{\delta}$ modes replacing $g$ does not fulfill
the sourceless Maxwell equations.For this reason
the replacement  $g \rightarrow
g_{\delta}$ is treated as a regularization which subsequently is to be removed.
The regularization (9) can be related to the one of our earlier paper
\cite{hab2} by means of the Laplace transform
\begin{displaymath}
exp(-\delta^{2}|{\bf x}|^{8})=\int_{0}^{\infty}ds h(s)exp(-s\sqrt{\delta}
|{\bf x}|^{2})
\end{displaymath}
where $h(s)$ is determined by the inverse Laplace transform (we made use
of $(iy)^{4}=y^{4}$ )
\begin{equation}
h(s)=\frac{1}{2\pi}\int_{-\infty}^{\infty}dy exp(-y^{4}+iys)
\end{equation}
It follows from eq.(10) that $h$ is a real, square integrable and bounded
function on $R$.

We obtain now the following  Feynman formula

{\bf Theorem 1}

Consider the Hamiltonian $H_{R}+H_{P}$ (eqs.(2)-(3)),where the
number of  modes is finite and $g(k,{\bf x})$
are replaced by $ g_{\delta}(k,{\bf x})$( this Hamiltonian is denoted
$H_{\delta}$).Assume that
$g_{\delta}(k,{\bf x})$
can be analytically extended to the region $V_{C}$

\begin{displaymath}
V_{C}=({\bf z}\in C^{3}:{\bf z}={\bf x} + (1+i){\bf y},{\bf x}\in V,{\bf
y}\in R^{3})
\end{displaymath}
and remain bounded analytic functions in $V_{C}$. Assume  that the
external potentials $u_{r} $ are
also analytic and bounded in $V_{C}$.Let us consider the initial state
of the form $\chi\Phi$ where $\chi$ is the ground state of $H_{R}$
and $\Phi$ is square integrable.
Then, the unitary semigroup $U^{\delta}_{t}$ ($t\geq 0$)
generating the Schr\"odinger time evolution has for a sufficiently
small time t the probabilistic
representation
\begin{equation}
\begin{array}{l}
(U^{\delta}_{t}\psi)({\bf A},{\bf x})\equiv
(exp(-iH_{\delta}\frac{t}{\hbar})\psi)({\bf A},{\bf x})=\chi({\bf A})
\cr
E\bigg[exp\Big(i\sum_{r}
\frac{e_{r}}{\hbar c}
\int_{0}^{t}{\cal A}_{\delta}(\tau,{\bf
x}_{r}+\lambda\sigma_{r}{\bf b}_{r}(\tau))\lambda\sigma_{r}d{\bf
b}_{r}(\tau)\Big)
\cr
exp\Big(-\frac{i}{\hbar}\sum_{r,s}e_{r}e_{s}\int_{0}^{t}\phi_{\delta}({\bf
x}_{r}+\lambda\sigma_{r}{\bf b}_{r}(\tau),{\bf
x}_{s}+\lambda\sigma_{s}{\bf b}_{s}(\tau))d\tau-
\cr
-\frac{i}{\hbar}\sum_{r}\int_{0}^{t}u_{r}({\bf
x}_{r}+\lambda\sigma_{r}{\bf b}_{r}(\tau))d\tau\Big)

\Phi({\cal A}_{\delta}(t,{\bf A}),{\bf x}+\lambda \sigma {\bf b}(t))\bigg]
\end{array}
\end{equation}
{\bf Remarks}

1.The index $\delta$ at the potentials ${\cal A}$ and $\phi$ means
that in their definitions we made the replacement $g \rightarrow
g_{\delta}$.

2.It does not matter whether the stochastic integral in eq.(10)
is in the Stratonovitch or Ito sense \cite{sim} because both integrals are
equal for transverse vector potentials.

3.If $V$ is a parallelepiped then $cos({\bf kx})$ and $sin({\bf kx})$
constitute the eigenmodes,hence $g_{\delta}(k,{\bf x})$
are analytic and bounded in $V_{C}$.

4. The wave function in the theorem describes the quantum
electromagnetic field interacting with the n-particle system.We denoted the
coordinates in the argument of $\psi$ and $\Phi$ collectively by ${\bf
x}$ and ${\bf b}$.

5. We may write $U_{t}=(U_{\frac{t}{n}})^{n}$, where the natural number
n is chosen large enough . In such a case we
can obtain a probabilistic representation for
 arbitrarily large time through a repetition of the formula (11).

6.We consider here only $t \geq 0$ the evolution for $ t \leq 0$ can
be obtained by time reflection as discussed in ref.\cite{hab2}.

{\bf Proof}: the proof is based on  general methods of the
construction of semigroups \cite{dyn}.The factor in the expectation
value in eq.(11) is an example of a non-anticipating multiplicative
functional. Such a multiplicative functional defines a semigroup.
There remains to prove that the multiplicative functional is integrable
(so that the expression (11) is finite) and that the generator of the
semigroup is $-\frac{i}{\hbar}H_{\delta} $.The integrability results from
our assumptions implying that $\phi_{\delta}$ and ${\cal A}_{\delta}$
are bounded.Then,for each ${\cal A}_{\delta}$
\begin{displaymath}
  exp\left(z\int_{0}^{t}{\cal A}_{\delta}(\tau,{\bf
x}+\lambda\sigma{\bf b}(\tau))\lambda\sigma d{\bf
b}(\tau)\right)
\end{displaymath}
is integrable for any complex $z$ on the basis of an estimate
proved in ref.\cite{car} and discussed in our earlier paper
\cite{hab2}. There remains the Gaussian integral over ${\cal A}_{\delta}$.
Owing to the $\delta$-regularization the covariance of ${\cal A}_{\delta}$
is a bounded continuous function. Hence, the ${\cal A}_{\delta}$-integral
is bounded by an exponential of a continuous bilinear form in the Brownian
motion ${\bf b}$ (this bilinear form is discussed in more detail in sec.5).
The expectation value of such an exponential is finite for a
sufficiently small time (see \cite{sim}\cite{hab2}). We can refer now to
the Fubini theorem in order to perform first the ${\cal A}_{\delta}$-integral
and to conclude that the expectation value in eq.(11) is finite.
Finally, the computation of the generator involves
the limit $t\rightarrow 0$. In such a case the calculations are reduced to
an elementary stochastic calculus \cite{sim} which does not depend
on the fact whether the potentials are real or complex. Let us
explain in more detail only the quantum electromagnetic part which
constitutes a new ingredient in comparison to ref.\cite{hab2}. In the
differentiation of ${\cal A}_{\delta}(t,{\bf x})$ over $t$ we
expand ${\cal A}_{\delta}$ in $q(t;k,\nu)$.Then, the differentiation
of $q$ is determined by eq.(5).As a result of the $q(t;k,\nu)$
differentiation we obtain the photon part of $H$
\begin{displaymath}
\chi^{-1}H_{R}\chi=-\sum_{k,\nu}
\left(\frac{\hbar^{2}}{2\mu(k)}\frac{\partial^{2}}
{\partial y(k,\nu) \partial y(k,\nu)} + \hbar\omega (k)y(k,\nu)\frac{\partial}
{\partial y(k,\nu)}\right)
\end{displaymath}
The remaining terms result from differentiation in time in the same
way as in the conventional form of the Feynman-Kac formula \cite{sim}.

Let us mention the imaginary time (Euclidean) version of eqs.(5)
and (11) (the Feynman-Kac formula).The Feynman-Kac formula
can be applied to more singular potentials. However, there are
stronger conditions on the possible decrease of the scalar potential
to minus infinity. The imaginary time version can be useful in a
study of the spectrum, but is unsuitable for an investigation of
dynamics.So, the imaginary time version of eq.(5) reads
\begin{equation}
dq(t;k,\nu)=-\omega(k) q(t;k,\nu)dt +  \sigma_{k} db(t;k,\nu)
\end{equation}
whereas the imaginary time version of eq.(11) takes the form
\begin{equation}
\begin{array}{l}
\left(exp(-H \frac{t}{\hbar})\psi\right )({\bf A},{\bf x})=
\cr
\chi({\bf A})
E\bigg[ exp\Big(i\sum_{r}\frac{e_{r}}{\hbar c}\int_{0}^{t}{\cal A}(\tau,{\bf
x}_{r}+\sigma_{r}{\bf b}_{r}(\tau))\sigma_{r}d{\bf
b}_{r}(\tau)\Big)
\cr
exp\Big(-\frac{1}{\hbar}\sum_{r,s}e_{r}e_{s}\int_{0}^{t}\phi({\bf
x}_{r}+\sigma_{r}{\bf b}_{r}(\tau),{\bf
x}_{s}+\sigma_{s}{\bf b}_{s}(\tau))d\tau-
\cr
-\frac{1}{\hbar}\sum_{r}\int_{0}^{t}u_{r}({\bf
x}_{r}+\sigma_{r}{\bf b}_{r}(\tau))d\tau\Big)
\Phi({\cal A}(t,{\bf A}),{\bf x}+ \sigma {\bf b}(t))\bigg]
\end{array}
\end{equation}
where ${\cal A}(t,{\bf x}) $ is expressed by $q(t;k,\nu)$ in the
same way as in the real time case (eq.(8)).Eq.(13) holds true under the
assumption that the number of modes is finite and $g(k,{\bf x})$ are
bounded functions on $R^{3}$ (in such case no $\delta$-regularization is
needed). It is easy to take
the limit of an infinite number of modes if the formfactors
$f(k)$ decay sufficiently fast for large k
(e.g.$|f(k)|\leq const |{\bf k}|^{-2}$).
\section{Probabilistic description of quantum dynamics at finite
temperature}
A mixed state can be described as a random pure state.Let us begin
with one degree of freedom.Assume that the (classical) probability
$\rho_{n}$ of a state $|n>$ is given.We introduce a random
variable $w$ such that
\begin{displaymath}
E[\overline{w}^{n}w^{m}]=\delta_{mn}n!\rho_{n}
\end{displaymath}
Then, the density matrix can be expressed in the form
\begin{equation}
\rho=E[|\chi><\chi|]
\end{equation}
where
\begin{equation}
|\chi>=\sum_{n=0}^{\infty}\frac{1}{\sqrt{n!}}w^{n}|n>
\end{equation}
Let us consider the quantum harmonic oscillator with the frequency
$\omega(k)$ at temperature T . If $|n>$ is the n-th excited
energy eigenstate
of the oscillator then for the Gibbs distribution
\begin{displaymath}
\rho_{n}=\left(1-exp(-\beta\hbar\omega(k))\right)exp(-\beta\hbar\omega(k)n)
\end{displaymath}
where $\beta=(KT)^{-1}$ and $K$ is the Boltzman constant.
It is easy to see that the expectation value (14) can be realized by
the Gaussian distribution
\begin{displaymath}
Tr(\rho F)=Tr\left(E[|\chi><\chi| F]\right)
=\left(1-exp(-\beta\hbar\omega(k))\right)
Tr\left(\int d\rho_{k}(w)
F|\chi><\chi|\right)
\end{displaymath}
where $d\rho_{k}$ is the Gaussian probability measure
\begin{displaymath}
d\rho_{k}(w)=\pi^{-1}dwd{\overline w}exp(\beta\hbar\omega(k))exp(
-exp(\beta\hbar\omega(k))|w|^{2})
\end{displaymath}
We can sum up the series (15) . In the coordinate representation we obtain
an expression for the coherent state
\begin{displaymath}
<y|\chi_{k}(z)>=(\frac{\eta_{k}}{\pi})^{\frac{1}{4}}
exp(-\frac{1}{2}\eta_{k}
y^{2}-\frac{\eta_{k}}{4} z^{2}+\eta_{k} zy)
\end{displaymath}
where
\begin{displaymath}
\eta_{k}=\frac{\hbar}{\mu(k)\omega(k)}
\end{displaymath}
and we defined
\begin{displaymath}
z=\sqrt{\frac{2}{\eta_{k}}}w
\end{displaymath}
$|\chi(z)>$ is just a coherent state with the mean value of the position
equal to $Rez$.Under the time evolution
\begin{equation}
\chi_{k}(t,y)\equiv<y|exp(-iH_{k}\frac{t}{\hbar})|\chi(z)>=
<y|\chi(exp(-i\omega(k)t)z)>
\end{equation}
According to our general scheme \cite{hab6} the time evolution of the
wave function $|\chi>$ (eq.(15)) determines the correlation functions
in the mixed state $\rho$. In more detail, we define first the
stochastic process ($0\leq \tau \leq t$)
\begin{equation}
dq(\tau)=\frac{i\hbar}{m}\chi(t-\tau,q)^{-1}\nabla \chi(t-\tau,q)d\tau +
\lambda\sigma db(\tau)
\end{equation}
Then, the correlation functions of Heisenberg operators can be expressed
by the solution $q(t,y)$ of the stochastic equation with the initial
condition $y$ e.g.
\begin{equation}
Tr(y(t)y(0)\rho)=E[\int dy|\chi_{t}(y)|^{2}yq_{t}(y)]
\end{equation}
In the case of the coherent state $\chi_{k}$ eq.(17) reads
\begin{equation}
dq(\tau)=-i\omega(k)qd\tau
+i\omega(k)zexp(-i\omega(k)(t-\tau))d\tau
+\lambda\sigma_{k}db(\tau)
\end{equation}
Its solution entering the formula (18) is
\begin{equation}
 \begin{array}{l}
q(t;y,z)=exp(-i\omega(k)t)y+iexp(-i\omega(k)t)zsin(\omega(k)t)+
\cr
+\lambda\sigma_{k}\int_{0}^{t}exp(-i\omega(k)(t-\tau))db(\tau)
\end{array}
\end{equation}
The correlation functions in eq.(18) can easily be computed from eqs.(16)
and (20)
\begin{equation}
\begin{array}{l}
Tr\Big(y(t)y(0)\rho\Big)=\Big(1-exp(-\beta\hbar\omega(k))\Big)
\int d\rho_{k}(w)dy
|<y|\chi_{k}(exp(-i\omega(k)t)w)>|^{2}
\cr
\Big(exp(-i\omega(k)t)y^{2}
+\frac{i\hbar\sqrt{2\eta_{k}}}{\mu(k)\omega(k)}sin(\omega(k)t)
exp(-i\omega (k)t)wy\Big)
\cr
=\frac{\hbar}{2\mu(k)\omega(k)}\bigg(cth(\frac{1}{2}\beta\hbar\omega(k))
cos(\omega(k)t)-isin(\omega(k)t)\bigg)
\end{array}
\end{equation}
According to the formulae (2) and (8) the free quantum electromagnetic
field is a sum of oscillators.The density matrix is a product of the
density matrices with different masses and frequencies.Hence, we obtain
directly from eqs.(8) and (21)
\begin{equation}
\begin{array}{l}
Tr\bigg(A_{j}(t,{\bf x})A_{l}(0,{\bf x}^{\prime})\rho\bigg)\equiv
G_{\beta}({\bf x},{\bf x}^{\prime};t)_{jl}=
\cr
\frac{2\pi\hbar c}{V}\sum_{k}\frac{f(k)^{2}}{|{\bf k}|}\delta^{tr}_{jl}
g(k,{\bf x})g(k,{\bf x}^{\prime})
\cr
\bigg(cth(\frac{1}{2}\beta \hbar c|{\bf k}|)
cos(c|{\bf k}|t) -isin(c|{\bf k}|t)\bigg)
\end{array}
\end{equation}
The expectation value in the ground state $\chi$ of the free
electromagnetic field in a
cavity is a special case of eq.(22) corresponding to the limit $\beta
\rightarrow \infty $ ($t \geq 0$)
\begin{equation}
\begin{array}{l}
D_{jl}({\bf x},{\bf x}^{\prime};t)\equiv
<\chi|A_{j}(t,{\bf x})A_{l}(0,{\bf x}^{\prime})|\chi>=
\cr
\frac{2\pi\hbar c}{V}\sum_{k}\frac{f(k)^{2}}{|{\bf k}|}\delta^{tr}_{jl}
g(k,{\bf x})g(k,{\bf x}^{\prime})exp(-ic|{\bf k}|t)
\end{array}
\end{equation}
The expression for correlation functions  of an
 n-particle system in equilibrium with the black body radiation
 at temperature T
described by the Hamiltonian $H_{R}+H_{P}$ (2)-(3) follows directly from
our description (14) of a mixed state as a random pure state. In this
way we obtain directly from Theorem 1
\begin{equation}
\begin{array}{l}
Tr\bigg(U_{t}^{+}\Phi_{1}({\bf A},{\bf x})U_{t}\Phi_{2}({\bf A},{\bf
x})\rho\bigg)=\int\prod_{k,\nu}dy(k,\nu)d{\bf x}\Phi_{1}({\bf A},{\bf x})
\cr
E\bigg[\Phi_{2}({\cal A}^{z}_{\delta}(t,{\bf A}),{\bf x}
+\lambda\sigma {\bf b}(t))
 exp\bigg(i\sum_{r}\frac{e_{r}}{\hbar c}\int_{0}^{t}{\cal A}^{z}_{\delta}
 (\tau,{\bf
x}_{r}+\lambda\sigma_{r}{\bf b}_{r}(\tau))\lambda\sigma_{r}d{\bf
b}_{r}(\tau)\bigg)
\cr
exp\bigg(-\frac{i}{\hbar}\sum_{r,s}e_{r}e_{s}\int_{0}^{t}\phi_{\delta}({\bf
x}_{r}+\lambda\sigma_{r}{\bf b}_{r}(\tau),{\bf
x}_{s}+\lambda\sigma_{s}{\bf b}_{s}(\tau))d\tau-
\cr
-\frac{i}{\hbar}\sum_{r}\int_{0}^{t}u_{r}({\bf
x}_{r}+\lambda\sigma_{r}{\bf b}_{r}(\tau))d\tau\bigg)\bigg]
\end{array}
\end{equation}
In eq.(24) ${\cal A}^{z}$ means that in the expression (8) for ${\cal
A}$ we insert the solution (20) depending on the coherent state
random variable $z_{k}$ as well as on the initial condition $y(k,\nu)$.
Then, in eq.(24) the average is taken over the Brownian motion,the
random variables $z_{k}$ as well as over the initial values $y(k,\nu)$
defining ${\bf A}$.Note, that the left hand side of eq.(24) is simple
if $\Phi_{1}=1$ whereas the right hand side is quite complicated.
This is so, because the unitarity condition applied at the left hand
side when expressed by a path integral takes a complicated form
(a condition of a time invariance of the expectation value is an analog
of unitarity).
\section{Quantum mechanics in QED environment}
In quantum mechanics we usually consider idealized systems consisting
of a finite number of particles. However, real systems are open and
interact in an uncontrollable way with an infinite environment. It
is believed that if the interaction with the environment is weak then
it is negligible. A model suggested by Caldeira and Leggett \cite{leg2}
indicates that this may be not the case concerning processes on the mesoscopic
scale. In particular, the disappearance of the quantum interference
in macroscopic systems may be a result of an interaction
with an environment (besides the Caldeira and Leggett papers see also
\cite{zur} \cite{gel} \cite{hab5}). However, the model of
an environment discussed in the above mentioned papers does not
come from any concrete physical environment ( the claims
in refs.\cite{ul}\cite{cal} that such an environment
can be considered as an approximation to QED is rather unconvincing). 
There appeared some
suggestions \cite{zeh} \cite{san} that QED (e.g. in the form of
 the black body radiation)
constitutes the environment enforcing the appearance of classical
properties of macroscopic systems. Unfortunately, the quantum
electrodynamics discussed in these papers has been considered in such a
rough approximation that a conclusion concerning its effect was rather
unfounded. Another reason for a  discussion of an effect of a black
body radiation in quantum mechanics is its possible substantial effect on
Rydberg atoms.

A restriction to observables associated with a system of particles
means an average over all states of the quantum radiation field.
Let us note first that the random electromagnetic field (8) consists of two
parts:the $\hbar$-independent part (which we call 
${\cal A}^{vac}$) depending on the
initial values $y(k,\nu)$
\begin{equation}
{\cal A}^{vac}(t,{\bf x})=\sum_{k,\nu}\sqrt{4\pi\mu(k)}
cf(k){\cal E}(k,\nu)exp(-i\omega(k)t)y(k,\nu)g(k,{\bf x})
\end{equation}
and the rest
\begin{displaymath}
{\cal A}^{Q}\equiv {\cal A}-{\cal A}^{vac}
\end{displaymath}
An easy computation gives
\begin{equation}
\begin{array}{l}
E[{\cal A}^{Q}_{j}(\tau,{\bf x}){\cal A}^{Q}_{l}(\tau^{\prime},{\bf
x}^{\prime})]=
\cr
\frac{2\pi\hbar c}{V}\sum_{k}\frac{f(k)^{2}}{|{\bf k}|}\delta^{tr}_{jl}
g(k,{\bf x})g(k,{\bf x}^{\prime})\bigg(exp(-ic|{\bf k}| |\tau^{\prime}-\tau|)
-exp(-ic|{\bf k}| (\tau^{\prime}+\tau))\bigg)
\cr
\equiv D_{jl}({\bf x},{\bf x}^{\prime};|\tau^{\prime}-\tau|)-
D^{vac}_{jl}({\bf x},{\bf x}^{\prime};\tau^{\prime}+\tau))
\end{array}
\end{equation}
where we divided the correlation function (26) into a time-translation
invariant part and the rest. If the $\delta$-regularization $g
\rightarrow g_{\delta}$ is inserted in eq.(26) then the corresponding
correlation function is denoted $D^{\delta}$.

We can calculate now transition amplitudes in quantum mechanics in a
QED environment.Let $\chi({\bf A})$ be the Fock vacuum for the radiation
field.Assume that $\Phi_{1}$ and $\Phi_{2}$ depend only on the particle
positions.Then the transition amplitudes can be expressed in the form
\begin{equation}
\begin{array}{l}
(\Phi_{1}\chi,U_{t}\Phi_{2}\chi)=                   
 \int d{\bf x}{\overline \Phi_{1}({\bf x})}E\bigg[\Phi_{2}({\bf x}+
 \lambda\sigma {\bf b}(t))
 \cr
 exp\bigg(-i\sum_{rs}\frac{e_{r}\sigma_{r}e_{s}\sigma_{s}}{2\hbar^{2}c^{2}}
 \int_{0}^{t}\int_{0}^{t}d{\bf b}_{r}(\tau)d{\bf b}_{s} (\tau^{\prime})
 D^{\delta}({\bf x}_{r}+\lambda\sigma_{r}{\bf b}_{r}(\tau),{\bf x}_{s} +
 \lambda\sigma_{s}{\bf b}_{s}(\tau^{\prime});|\tau-\tau^{\prime}|)\bigg)
 \cr
exp\bigg(-\frac{i}{\hbar}\sum_{r,s}e_{r}e_{s}\int_{0}^{t}\phi_{\delta}({\bf
x}_{r}+\lambda\sigma_{r}{\bf b}_{r}(\tau),{\bf
x}_{s}+\lambda\sigma_{s}{\bf b}_{s}(\tau))d\tau -
\cr
-\frac{i}{\hbar}\sum_{r}\int_{0}^{t}u_{r}({\bf
x}_{r}+\lambda\sigma_{r}{\bf b}_{r}(\tau))d\tau\bigg)\bigg]
\end{array}
\end{equation}
where $d{\bf b} d{\bf b}D$ means a sum over vector indices of
$D$ and vector indices of the Brownian motions of
individual particles.

{\bf Remarks}:

1. The conditions for applicability of the Fubini theorem are
fulfilled in the regularized theory for a sufficiently
small time (cp.Theorem 1). Double stochastic integrals
are discussed in refs.\cite{nel}\cite{ber}.

2. It is remarkable that the "vacuum part" in 
the exponential of the expectation
value (27) is absent. The "vacuum part" of eq.(26) canceled
with the contribution from the expectation value of ${\cal A}^{vac}$
and what remains is the time translation
invariant part of the expression (26).

The corresponding formula at finite temperature reads
\begin{equation}
\begin{array}{l}
Tr\bigg(U_{t}^{+}\Phi_{1}({\bf x})U_{t}\Phi_{2}({\bf x})\rho\bigg)=
\int d{\bf x}
\Phi_{1}({\bf x})E\bigg[\Phi_{2}({\bf x}+\lambda \sigma{\bf b}(t))
\cr
exp\bigg(-i\sum_{rs}\frac{e_{r}\sigma_{r}e_{s}\sigma_{s}}
{2\hbar^{2}c^{2}}
 \int_{0}^{t}\int_{0}^{t}d{\bf b}_{r}(\tau)d{\bf b}_{s} (\tau^{\prime})
 G^{\delta}_{\beta}({\bf x}_{r}+\lambda\sigma_{r}{\bf b}_{r}(\tau),
 {\bf x}_{s} +
 \lambda\sigma_{s}{\bf b}_{s}(\tau^{\prime});|\tau-\tau^{\prime}|)\bigg)
 \cr
exp\bigg(-\frac{i}{\hbar}\sum_{r,s}e_{r}e_{s}\int_{0}^{t}\phi_{\delta}({\bf
x}_{r}+\lambda\sigma_{r}{\bf b}_{r}(\tau),{\bf
x}_{s}+\lambda\sigma_{s}{\bf b}_{s}(\tau))d\tau-
\cr
-\frac{i}{\hbar}\sum_{r}\int_{0}^{t}u_{r}({\bf
x}_{r}+\lambda\sigma_{r}{\bf b}_{r}(\tau))d\tau\bigg)\bigg]
\end{array}
\end{equation}
where $G^{\delta}_{\beta}$ is the finite temperature correlation
function (22) with $g \rightarrow g_{\delta}$.

We can see from eq.(27) that the quantum electromagnetic environment
contributes an additional non-local interaction (described by $D$ )
 between particles of the system.
One could say that the particles interact through an exchange of photons
from the environment.  As $D$ (at zero temperature) is proportional
 to $\hbar$
 the additional interaction is of the order
of the fine structure constant $\alpha$. From eq.(22) we can see that at large
temperature the interaction becomes proportional to the temperature.
However, because of the light velocity in the denominator, this
interaction seems  negligible at the room temperatures. Nevertheless,
it should be realized that such
a comparison of the strength of interactions makes sense only for
interactions of the same type. The interaction induced by the QED
environment is of different type than the Coulomb interaction $\phi$
and the external potential $u$.It depends on velocities and is
non-local. Such interactions may be relevant on the large time
and space scale.

\section{ Removal of cutoffs and regularizations}
We restricted ourselves till now to a finite number of modes (an
ultraviolet cutoff) in a finite volume V and we introduced additionally
the $\delta$-regularization which is demanded by the functional
integral representation in the real time (it would be superfluous
in the Feynman-Kac formula).Together with a finite number of
modes we considered also the formfactor $f(k)$ which can describe
a finite distribution of the charge.We can allow the number of modes
to increase to infinity if the formfactor decreases sufficiently fast
(in $k$).In such a case the formfactor fulfills the role of the
ultraviolet cutoff. We restrict ourselves from now on to the case
when $V$ is a cube. Then, $g(k,{\bf x})$ consist of pairs $(cos({\bf
kx}),sin({\bf kx}))$ (with a certain restriction on ${\bf k}$
ensuring that the basis is orthogonal,see \cite{kro}\cite{lan}).
Hence,
\begin{equation}
g(k,{\bf x})g(k,{\bf x}^{\prime})\rightarrow cos({\bf kx})cos({\bf
kx}^{\prime})+  sin({\bf kx})sin({\bf kx}^{\prime})=cos({\bf kx-kx}^{\prime})
\end{equation}
It can be seen from eqs.(26)-(27) that in a cube (with
periodic boundary conditions) the interaction is invariant under
space and time translations. Moreover, if the temperature is equal
to zero, the cube tends to $R^{3}$ and the ultraviolet regularization
is removed ($f(k)=1$) then  the interaction becomes
relativistic invariant ( in such a case $D$ coincides with the
conventional $D_{+}$).It is well-known that in the relativistic
theory $D_{+}$ is singular on the light cone. 
The origin of this sigularity can be
seen from eqs.(26) and (29); the oscillating factor becomes a constant
on the light cone,hence the sum over wave vectors tends to infinity.
In such a case the double stochastic integral in eq.(27) is infinite.
The ultraviolet problem is unavoidable in QED and presumably can only be
solved perturbatively through an expansion of the exponential
 in power series (which is a power series in
$\alpha$). We consider in this section the remaining regularizations.

{\bf Theorem 2}

Let V be a cube . Assume that the number of modes is finite . Assume
also that the external potentials $u_{r}$ are bounded on $V_{C}$.
 Then, the limit
 $\delta \rightarrow 0 $ of
$exp\left(-iH_{\delta}\frac{t}{\hbar}\right)\psi({\bf A},{\bf x})$ exists for
each ${\bf A}$ and ${\bf x}$ and sufficiently small $t$.
 The limit $\delta \rightarrow 0$ defines a unitary  evolution
solving the Schr\"odinger equation
\begin{displaymath}
i\hbar\partial_{t}\psi_{t}=(H_{R}+H_{P})\psi_{t}
\end{displaymath}
with the initial condition $\psi$. As a consequence the limits $\delta
\rightarrow 0$ of the transition amplitude (27) and the finite temperature
 correlation
functions (28) also exist for a sufficiently small time.

{\bf Proof}: We apply  the representation (9)-(10) to the $\delta$-
regularization. We expand the exponentials in eq.(11) in power series in
e.Then, it follows from the results
of ref.\cite{hab2} (Theorem 3) and from the Lebesgue dominated convergence
 theorem (because ${\cal A}_{\delta}$ and $\phi_{\delta}$ are bounded)
 that
for each ${\cal A}^{Q}$ the limit $\delta \rightarrow 0$ is exchangeable
with the sum.There remains the Gaussian integral over ${\cal A}^{Q}$ in
the power series. From  estimates of the combinatorial factor
in front of the term $({\cal A}^{Q})^{n}$ done in \cite{hab2} (based on
ref.\cite{car}) it follows that it is bounded by $\frac{1}{\sqrt{n!}}$.
Hence, after the Gaussian integration  of the term $({\cal A}^{Q})^{n}$
this term becomes bounded by $c(t)^{n}$ where $c(t)$ is proportional to t
because the upper limit in the stochastic integration is equal to t.
Hence, if t is sufficiently small then the power series is bounded by
 a convergent geometric series. The proof of the convergence of the
 power series for the transition amplitude (27) and the correlation
 functions (28) is similar. In the proof we replace the Gaussian integral
  with the covariance (26) by the Gaussian integral with the covariance
  (23).

{\bf Corollary 3}

Assume that there exist $a>0$ and $R(a)>0$ such that
\begin{equation}
|f(k)|\leq R(a)exp(-a|{\bf k}|^{2})
\end{equation}

i)Let N be the number of modes. Then, the limit $N \rightarrow \infty$
of the transition amplitude (27) and the finite temperature correlation
functions (28) exists uniformly in $\delta$ for a sufficiently small time t.
The limits $N \rightarrow \infty$ and $\delta \rightarrow 0$ are
exchangeable and can be taken simultaneously.

ii)If time t is sufficiently small then after $N \rightarrow \infty$
 the limit $V \rightarrow R^{3}$
of the transition amplitude and the same limit of 
the finite temperature correlation
functions can be performed according to the rule
\begin{displaymath}
\frac{1}{V}\sum_{k} \rightarrow \frac{1}{(2\pi)^{3}}\int d^{3}k
\end{displaymath}
The limit $V \rightarrow R^{3}$ is uniform in $\delta$. Hence,the
limits $V \rightarrow R^{3}$ and $\delta \rightarrow 0$ are
exchangeable.

{\bf Remark}: We could define directly the evolution operator $U_{t}$
(11) for an infinite number of modes and a sufficiently small time by the
 same argument which we apply in the proof below.
 In principle, an operator can be recovered from the transition amplitude.

{\bf Proof}:It is sufficient to note that with the bound on the
formfactor $f(k)$ the limits considered in the Corollary can be
taken at each order of the perturbation series in e. This is so, because
according to the results of ref.\cite{hab2} (proven under a condition
equivalent to the bound on the formfactor assumed here) and Theorem 2 the
perturbation series is convergent.Then , the existence of the limit
$N\rightarrow \infty$ is easy to check at each order of the perturbation
expansion.The infinite volume limit and the convergence of the sum over
$k$ to an integral is a consequence of Riemann$^{\prime}$s definition of
an integral.

We approached QED starting from a well-defined Hamiltonian consisting
of a finite number of modes of the quantum electromagnetic field
in a finite volume. This is a model often studied in quantum optics.
However, we could consider the final formula (27) where the limits
$\delta \rightarrow 0$ ,$N \rightarrow \infty$ and $V \rightarrow R^{3}$ are
performed directly under the expectation value on the right hand side
 in the exponential factor. We would obtain an expression defined by
  the stochastic integrals of $D$ and $\phi$.
 This could be considered as a definition of QED.
 In fact, if we first integrated over ${\cal A}$ in the (imaginary time)
 Feynman-Kac formula and afterwards performed in a formal way the
 analytic continuation on the path space, then we would arrive at
 the formula (27). In Theorem 2 and Corollary 3
  we have shown that the limits on the
 right hand side exist for a sufficiently small time
 and define a limit of an evolution operator. If the limits are
performed under the expectation value then  the expectation value
may be infinite or even if it is finite we may lose a connection
with the Hamiltonian formalism. In the rest of this section we
investigate the double stochastic integral of the formula (27)
treating the right hand side  as a definition of the Feynman integral
representation of the left hand side. We emphasize the
following result of ref.\cite{ber}

{\bf Lemma 4}

The sufficient conditions required for a definition of the double stochastic
integral in eq.(27) as an almost surely finite random variable read
\begin{equation}
E\left[\int_{0}^{t}d\tau\int_{0}^{\tau}d\tau^{\prime}|D\left({\bf b}(\tau),
{\bf b}(\tau^{\prime});
\tau -\tau^{\prime}\right)|^{2}\right] < \infty
\end{equation}
and
\begin{displaymath}
E\left[\int_{0}^{t}d\tau |D({\bf b}(\tau),{\bf b}(\tau);0)|^{2}\right]
< \infty
\end{displaymath}
Applying Lemma 4 we can show

{\bf Lemma 5}

Under the assumption (30) the double stochastic integral 
is defined as an almost
surely finite random variable.

{\bf Proof}: Under our assumptions the correlation function $D$ for
equal times is a finite constant. There remains to prove that the
integral (31) is finite.The expectation value (31)is equal to
\begin{displaymath}
\begin{array}{l}
\int_{0}^{t}d\tau\int_{0}^{\tau}d\tau^{\prime}d^{3}x d^{3}yd^{3}k
 d^{3}k^{\prime}
 (2\pi)^{-3}
(\tau^{\prime}(\tau-\tau^{\prime}))^{-\frac{3}{2}}
\cr
exp(-\frac{{\bf x}^{2}}{2\tau^{\prime}}
-\frac{({\bf y}-{\bf x})^{2}}{2(\tau-\tau^{\prime})})
(|{\bf k}| |{\bf k}^{\prime}|)^{-1}
\cr
exp\left(-ic(|{\bf k}| +|{\bf k}^{\prime}|)(\tau -\tau^{\prime}) +
\lambda\sigma({\bf ky}-{\bf kx})+\overline{\lambda}
\sigma({\bf k^{\prime}y}-{\bf k^{\prime}x}) \right)
\end{array}
\end{displaymath}
We change variables $({\bf x},{\bf y})\rightarrow ({\bf x},{\bf z})$ where
\begin{displaymath}
{\bf x}-{\bf y}=\sqrt{\tau -\tau^{\prime}}{\bf z}
\end{displaymath}
After such a change of variables and a calculation of the ${\bf x}$-integral
the integrand becomes regular in $\tau$ and $\tau^{\prime}$ whereas
the integrability at large $k$ and $k^{\prime}$is ensured by the
assumption (30).

Finally, we consider an explicit example
\begin{equation}
f(k)=exp(-\frac{a}{2} |{\bf k}|)
\end{equation}
(where $a>0$) which does not fulfill the assumption (30).
Nevertheless, we can show

{\bf Proposition 6}

Let us restrict ourselves in Theorem 1 to a single particle.

i)If $f(k)$ is defined by eq.(32) then the function D (the case of
an infinite volume) has
 an analytic continuation to the region $V_{C}(0)=({\bf z}\in C^{3}:
 {\bf z}=(1+i){\bf y},{\bf y}\in R^{3} )$. Moreover, D is bounded
 in this region.

ii) Assume that the external potential u is bounded by a quadratic form
in the region $V_{C}$ of Theorem 1. Then, the functional integral (27)
with the formfactor (32) is finite for a sufficiently small time.

{\bf Remark}: The multiparticle case  with the formfactor (32) is
more involved, because D and $\phi$ are  meromorphic multivalued
functions in $V_{C}$.

{\bf Proof}: A direct calculation from eq.(23) (see also eq.(26))
gives
\begin{displaymath}
D_{jl}({\bf x},{\bf
x}^{\prime};\tau-\tau^{\prime})=\frac{\delta^{tr}_{jl}}{4\pi}
\left((a+ic|\tau-\tau^{\prime}|)^{2}+({\bf x}-{\bf x}^{\prime}\right)^{2})^{-1}
\end{displaymath}
For a single particle the function D in the double stochastic integral
of eq.(27) takes the form
\begin{displaymath}
\begin{array}{l}
D_{jl}(\lambda\sigma{\bf b}(\tau),\lambda \sigma{\bf b}(\tau^{\prime})
;\tau-\tau^{\prime})\equiv \delta^{tr}_{jl} D=
\cr
\frac{1}{4\pi} \delta^{tr}_{jl}
\left(a^{2}-c^{2}(\tau-\tau^{\prime})^{2}+2iac|\tau-\tau^{\prime}|+i\sigma^{2}
({\bf b}(\tau)-{\bf b}(\tau^{\prime}))^{2}\right)^{-1}
\end{array}
\end{displaymath}
$\delta^{tr}$ is a bounded operator. So, it is sufficient to discuss
the scalar function D behind $\delta^{tr}$ ( a more detailed
demonstration would involve an application of the Ito formula
;we  express the stochastic integral with respect to the
longintudinal modes by an ordinary integral which would be bounded
owing to the regularization). It is easy to see that D is bounded
because, if
\begin{displaymath}
|\tau -\tau^{\prime}|c\leq\frac{a}{2}
\end{displaymath}
then  $|D|\leq (3\pi a^{2})^{-1}$ because the real part of the
denominator of D is different from zero. If
\begin{displaymath}
|\tau -\tau^{\prime}|c\geq\frac{a}{2}
\end{displaymath}
then $|D|\leq (3\pi a^{2})^{-1}$, because the imaginary part of
the denominator of D is different from zero.

When D is a bounded continuous function and u is quadratically  bounded
then the integrand (27) can be bounded by the Gaussian integral. In such
a case the well-known theorems (see e.g.\cite{sim};
{\bf b} can be expressed by independent Gaussian random variables) show
the integrability proving ii).

Even if D is bounded it may be not easy to prove integrability in
eq.(27) for arbitrarily large time. We have to show that the
negative part of D decreases sufficiently fast when $ |\tau
-\tau^{\prime}|$ grows. If this is not the case we have to continue
the formula (27) beyond the time-interval where it loses its meaning.
We encounter this problem already for a harmonic oscillator. A solution
of the problem is  well-known in this case . For an anharmonic
oscillator it is discussed in a semiclassical approximation
(Maslov indices).

\section{The Feynman kernel of the evolution operator}
We wish to derive an explicit Feynman integral formula
for the evolution kernel of the operator (11). We mainly discuss
 in this section a finite number of modes.
However, first we would like to explain that in principle
there is no difficulty in defining an evolution kernel in infinite
number of dimensions . There is no Lebesgue measure 
in infinite number of dimensions. For this
 reason we define a kernel K as a transformation of a set of measures
$ {\cal G} $ into a set of functions $L^{2}(d\mu)$ integrable with
respect to a certain measure $\mu$. In QED we choose for $\mu$ a
measure defined on vector fields which gives a realization of the Fock space
 of the transversal electromagnetic free field i.e. $\mu$
is the Gaussian measure with the covariance
\begin{displaymath}
\hbar\delta^{tr}_{jl}(-\triangle)^{-1}
\end{displaymath}
If $\Phi_{1}$ and $\Phi_{2}$ belong to $ L^{2}(d\mu)$ then we define
the kernel $K_{t}$ of the unitary evolution $U_{t}$ by the formula
\begin{displaymath}
\begin{array}{l}
(\Phi_{1},U_{t}\Phi_{2})=(\Phi_{1},K_{t}\Phi_{2} d\mu)\equiv
\int d^{3} x d\mu({\bf C}) \overline{\Phi_{1}}({\bf C},{\bf x})
\cr
 K(t,{\bf C},{\bf x};0,{\bf B},
 {\bf x}^{\prime})\Phi_{2}({\bf B},x^{\prime})d\mu({\bf B})d^{3}x^{\prime}
\end{array}
\end{displaymath}

We have obtained
a Feynman formula for the kernel of a quantum mechanical Hamiltonian
evolution in the case without the quantum electromagnetic field
in ref.\cite{hab3}. In order to extend this formula to a quantum
electromagnetic field let us start with a definition of the Gaussian
process $0\leq \tau \leq t$
\begin{equation}
Q^{t}(\tau;k,\nu)=\alpha^{t}(\tau;k,\nu)
+ \lambda\sigma_{k}\gamma^{t}(\tau;k,\nu)
\end{equation}
where $\alpha^{t}$ (we shall sometimes conceal some indices which
are inessential at the moment) is the solution of the oscillator equation
\begin{equation}
(\frac{\partial^{2}}{\partial\tau^{2}} +\omega(k)^{2})\alpha^{t}(\tau)=0
\end{equation}
with the boundary conditions $\alpha^{t}(0)=X$ 
and $\alpha^{t}(t)=Y$, explicitly
\begin{displaymath}
\alpha^{t}(\tau)= \frac{sin(\omega(k)(t-\tau))}{sin(\omega(k)t)}X +
\frac{sin(\omega(k)\tau)}{sin(\omega(k)t)}Y
\end{displaymath}
$\gamma^{t}$ is the Gaussian process fulfilling the zero boundary
conditions $\gamma^{t}(0)=\gamma^{t}(t)=0$ with mean zero and the covariance
\begin{equation}
E[\gamma^{t}(\tau)\gamma^{t}(\tau^{\prime})]={\cal
M}(\tau,\tau^{\prime})
\end{equation}
where
\begin{equation}
{\cal L}_{\tau}{\cal M}(\tau,\tau^{\prime})
\equiv(-\frac{\partial^{2}}
{\partial \tau^{2}}-\omega(k)^{2}){\cal M}(\tau,\tau^{\prime})=\delta(
\tau-\tau^{\prime})
\end{equation}
i.e.${\cal L}{\cal M}=1$.${\cal M}$ is the Green function of the
operator ${\cal L}$ with the  Dirichlet boundary conditions on the
interval [0,t].For $ t< \pi \omega(k)$ the kernel ${\cal M} $ is positive
definite, because for such a t the operator ${\cal L}$ is
 positive definite. This is so,because
the lowest eigenvalue of $-\partial^{2}$ is $(\frac{\pi}{t})^{2}$.
 In such a
case the random variable $\gamma^{t}$ is real. The explicit formula for
${\cal M} $ reads
\begin{displaymath}
\begin{array}{l}
{\cal M}(\tau,\tau^{\prime})=\frac{1}{\omega(k)}
\theta(
\tau^{\prime}-\tau)\bigg(sin(\omega(k)
(t-\tau^{\prime}))cos(\omega(k)(t-\tau)) -
\cr
- ctg(\omega(k)t)sin(\omega(k)(t-\tau))sin(\omega(k)(t-\tau^{\prime})\bigg) +
\cr
+\frac{1}{\omega(k)}
\theta(\tau-\tau^{\prime})
\bigg(sin(\omega(k)(t-\tau))cos(\omega(k)(t-\tau^{\prime}))-
\cr
- ctg(\omega(k)t)sin(\omega(k)(t-\tau))sin(\omega(k)(t-\tau^{\prime})\bigg)
\end{array}
\end{displaymath}
where $\theta$ is the Heaviside step function.

We can express the process $\gamma^{t}$ by the Brownian motion
\begin{equation}
\gamma^{t}(\tau)=sin(\omega(k)(t-\tau))\int_{0}^{\tau}
\Big(sin(\omega(k)(t-u))\Big)^{-1}db(u)
\end{equation}
We insert $Q^{t}$ (33) into the expansion (8) of the electromagnetic field
${\cal A}$ into eigenmodes. We obtain
\begin{equation}
{\cal A}(\tau,{\bf x};t,{\bf B},{\bf C})
={\cal A}^{cl}(\tau,{\bf x};t,{\bf B},{\bf
C})+ \lambda{\cal H}^{t}(\tau,{\bf x})
\end{equation}
where ${\cal A}^{cl}$ is the solution of  the wave equation
\begin{displaymath}
(\frac{1}{c^{2}}\frac{\partial^{2}}{\partial\tau^{2}}-\triangle){\cal
A}^{cl}=0
\end{displaymath}
with the two-point boundary condition ${\cal A}^{cl}(0,{\bf x};t,{\bf B},
{\bf C})={\bf B}({\bf x})$ and ${\cal A}^{cl}(t,{\bf x};t,{\bf B},{\bf
C})={\bf C}({\bf x} )$. ${\cal H}^{t}$ is a quantum fluctuation. Note that
the two-point boundary value problem (34) has the unique solution
 only if $t<
\frac{\pi}{\omega(k)}$ for every k.In terms of the initial condition
of eq.(34)
${\bf B} $ has the form
\begin{displaymath}
{\bf B}=\sum_{k,\nu}\sqrt{4\pi\mu(k)}cf(k){\cal E}(k,\nu)X(k,\nu)
g(k,{\bf x})
\end{displaymath}
We have a similar formula for ${\bf C}$ in terms of $Y(k,\nu)$.
The decomposition of the quantum field ${\cal A}$
(38) into a solution of a classical boundary value problem and a random
fluctuation resembles Boyer $^{\prime}$s stochastic electrodynamics \cite{boy}.
However, our random fluctuation ${\cal H}$ is multiplied by a complex
number $\lambda$.

The construction of the random electromagnetic field (38) fulfilling the
two-point boundary condition is a generalization of the construction
of the Brownian bridge $v$ \cite{sim}. This is the Gaussian
process defined on the interval [0,1] with zero two-point boundary
condition and the covariance
\begin{displaymath}
E[v_{j}(\tau)v_{l}(\tau^{\prime})]
=\delta_{jl}\Big(\theta(\tau-\tau^{\prime})\tau^{\prime}
(1-\tau)+ \theta(\tau^{\prime}-\tau)\tau(1-\tau^{\prime})\Big)
\end{displaymath}
The bridge connecting the points ${\bf x}$ and ${\bf x}^{\prime}$
which we apply in the Feynman formula has the form
\begin{equation}
{\bf p}(\tau;t,{\bf x},{\bf x}^{\prime})={\bf
x}^{\prime}\frac{\tau}{t}+\frac{t-\tau}{t}{\bf x}+
\lambda\sigma\sqrt{t}{\bf v}(\frac{\tau}{t})
\end{equation}
We can now express the evolution kernel by the bridge

{\bf Theorem 7}

i) Let $0 \leq t < min_{k}(\frac{\pi}{\omega(k)})$ then
the integral kernel
of the unitary semigroup $U_{t}$ of Theorem 1 (under the
assumptions of this theorem ) has the representation
\begin{equation}
\begin{array}{l}
K^{\delta}(t,{\bf C},{\bf x};0,{\bf B},{\bf x^{\prime}})=
K^{free}(t,{\bf C},{\bf x};0,{\bf B},{\bf x^{\prime}})
\cr
E\bigg[exp\bigg(i\sum_{r}\frac{e_{r}}{\hbar c}\int_{0}^{t}{\cal A}_{\delta}
(\tau,{\bf p}_{r}(\tau);t,{\bf B},{\bf C})
d{\bf p}_{r}(\tau)\bigg)
\cr
exp\bigg(-\frac{i}{\hbar}\sum_{r,s}e_{r}e_{s}\int_{0}^{t}\phi_{\delta}(
{\bf p}_{r}(\tau),{\bf p}_{s}(\tau))d\tau
-\frac{i}{\hbar}\sum_{r}\int_{0}^{t}u_{r}({\bf p}_{r}(\tau))d\tau\bigg)\bigg]
\cr
\equiv K^{free}{\cal R}
\end{array}
\end{equation}
where $K^{free}$ is the Feynman kernel of $exp(-iH_{R}\frac{t}{\hbar})$
\begin{displaymath}
\begin{array}{l}
K^{free}(t,{\bf C},{\bf x};0,{\bf B},{\bf x^{\prime}})=
\prod_{k}(2\pi i\sigma_{k}^{2} sin(\omega(k)t))^{-1}
\cr
\prod_{k,\nu}exp\bigg(\frac{i}{2\sigma_{k}^{2}}\Big(X(k,\nu) ctg(c|{\bf k}|t))
X(k,\nu)+ Y(k,\nu)ctg(c|{\bf k}|t))Y(k,\nu) -
\cr
      -2X(k,\nu)sin(c|{\bf k}|t))^{-1}Y(k,\nu)\Big)\bigg)
\end{array}
\end{displaymath}

ii) assume that i) holds true then under the assumptions of Theorem 2
 there exists the limit
$\delta \rightarrow 0$ of $K^{\delta}(t,{\bf C},{\bf x};
0,{\bf B},{\bf x^{\prime}})$
 for each ${\bf x}, {\bf x^{\prime}},{\bf B}$ and $ {\bf C}$
 if t is sufficiently small.

iii) the limit $V\rightarrow R^{3}$ exists for a sufficiently small time
if $f(k)=0$ for $|{\bf k}|>{\kappa}$ with a certain ultraviolet
cutoff ${\kappa}$

{\bf Remarks}

1.The allowed time in ii) and iii) may be smaller than in i) because we
additionally require a convergence of the perturbation series in powers
of the electric charge.

2.The formula for arbitrary time can in principle be derived by
a composition of kernels for a small time.

{\bf Proof}:In i) we can repeat the argument of ref.\cite{hab3}
in order to represent the kernel of the evolution operator of
Theorem 1 (eq.(11)) in terms of the corresponding bridge joining the initial
and the final points. Here, for the electromagnetic field we
apply the bridge (33) corresponding to the harmonic oscillator instead
the $v$-bridge corresponding to a free particle used in our earlier
paper \cite{hab3} (the imaginary time version of the oscillator
bridge is discussed in \cite{simao}).In part ii) we repeat the
proof of Theorem 2 showing that the perturbation series is convergent
uniformly in $\delta$ and that the $\delta$-regularization can be
removed term by term. In part iii) we apply the same argument as in
ii) of Corollary 3 to show that a sum over wave vectors ${\bf k}$ converges
to an integral when $V\rightarrow 0$.This is possible only if the
denominator in the bridge $Q^{t}$ (33) is different from zero. For this
reason we assumed the sharp ultraviolet cutoff in iii).

It seems that eq.(39) remains true for any $t\neq \frac{\pi}{\omega(k)}$.
A rough argument is as follows: it could be proved
in an independent way under our assumptions that the evolution
kernel of $H_{R}+ H_{P}$ has a convergent expansion around
the evolution kernel of $H_{R}$ (see ref.\cite{weistein} for
a closely related theorem). On the other hand eq.(40) defines such an
expansion for any $t\neq \frac{\pi}{\omega(k)}$. Hence, it is
sufficient to compare the expansion coefficients. These, expansion
coefficients must be the same because both kernels fulfil the
same equation. Another argument which could be applied comes from
the fact that an imaginary time version of eq.(40)(cp.eq.(13)) holds true
for arbitrarily large time. For an operator bounded from below
an analytic continuation in time exists (see a discussion in ref.
\cite{fel}). At points where the right hand side of eq.(40) is regular
it  must coincide with the integral kernel of the analytic continuation
of the semigroup $exp(-H\frac{t}{\hbar})$.

    \section{The semiclassical expansion}
A non-perturbative construction of QED is necessary
for a demonstration of a classical limit of QED. 
This is so, because the conventional
perturbative QED involves an expansion in the fine structure constant, where
$\alpha \rightarrow \infty$ when $\hbar\rightarrow 0$ ( see a discussion
of some problems concerning this limit in ref.\cite{bia}). We could obtain
the conventional perturbation expansion expanding eq.(27) in powers
of the electric charge (see \cite{fey1} for a relativistic
 generalization of eq.(27)). In this section we obtain a quasiclassical
form of the Feynman kernel (40) through an expansion in $\hbar$
(the semiclassical expansion of the Feynman kernel in
quantum mechanics with a rigorous version of the Feynman integral
is discussed in \cite{alb}\cite{aze}\cite{hab3}; for a version
without the Feynman integral see \cite{fuj}).
In such an approach the role of the two-point boundary value problem
and its quantum correction (38) become visible. After a derivation
of the quasiclassical form of the Feynman  kernel other versions of
the classical limit can be obtained through an expression of the
time evolution by the Feynman kernel. In particular, one can see more
explicitly the difference between QED and Boyer$^{\prime}$s stochastic
electrodynamics \cite{boy}. In the derivation of the semiclassical
expansion we follow our earlier paper \cite{hab3}.The first step is
a shift of variables $\gamma\rightarrow \gamma +\zeta\beta$ and
 ${\bf v}\rightarrow {\bf v}
 +z{\bf w}$, where ${\bf v}$ and ${\bf w}$ are real, whereas
  $z$ and $\zeta$ are complex numbers

{\bf Lemma 8}

Let $\beta(\tau)$ and ${\bf w}(\tau)$ be differentiable and their derivatives
square integrable on the intervals
$[0,1]$ and $[0,t]$ respectively.Moreover, assume that $\beta$ and $ {\bf w}$
are equal to zero on the ends of the respective intervals. Then,
with the assumptions and notation of i) 
of Theorem 7 we have the following identity
\begin{equation}
\begin{array}{l}
{\cal R}=
E\bigg[exp\bigg(-\frac{1}{2}\int_{0}^{t}
\sum_{k,\nu}\zeta(k,\nu)^{2}\beta(\tau;k,\nu){\cal L}
\beta(\tau;k,\nu)d\tau-
\cr
      -\int_{0}^{t}\sum_{k,\nu}\zeta(k,\nu){\cal L}
      \beta(\tau;k,\nu)\gamma(\tau;k,\nu)d\tau
     -\int_{0}^{1}\sum_{r}z_{r}\frac{d{\bf w}_{r}}{d\tau}d{\bf v}_{r}
     (\tau)-
  \frac{1}{2}\int_{0}^{1}\sum_{r}z_{r}^{2}\frac{d{\bf w}_{r}}{d\tau}
     \frac{d{\bf w}_{r}}{d\tau}d\tau +
\cr
+i\sum_{r}\frac{e_{r}}{\hbar c}\int_{0}^{t}{\cal A}_{\delta}
(\tau,{\bf p}_{r}(\tau)
+\lambda\sigma_{r}z_{r}{\bf w}_{r}(\tau);t,{\bf B}+{\bf F},{\bf C})
d({\bf p}_{r}(\tau)+\lambda\sigma_{r} z_{r}{\bf w}_{r}(\tau))\bigg)
\cr
exp\bigg(-\frac{i}{\hbar}\sum_{r,s}e_{r}e_{s}\int_{0}^{t}\phi_{\delta}(
{\bf p}_{r}(\tau)+\lambda\sigma_{r}z_{r}{\bf w}_{r}(\tau),
{\bf p}_{s}(\tau)+\lambda\sigma_{r} z_{r}{\bf w}_{r}(\tau))d\tau
\cr
-\frac{i}{\hbar}\sum_{r}\int_{0}^{t}u_{r}({\bf p}_{r}(\tau)
+\lambda\sigma_{r}z_{r}{\bf w}_{r})d\tau\bigg)\bigg]
\end{array}
\end{equation}
where ${\cal L}$ is defined in eq.(36) and
\begin{displaymath}
{\cal A}_{\delta}(\tau,{\bf x};t,{\bf B}+{\bf F},{\bf C})
\equiv{\cal A}_{\delta}(\tau,{\bf x};t, 
{\bf B},{\bf C})+{\cal F}_{\delta}(\tau,{\bf x})
\end{displaymath}
with
\begin{displaymath}
{\cal F}_{\delta}(\tau,{\bf x})
=\sum_{k,\nu}\sqrt{4\pi\mu(k)}cf(k){\cal E}(k,\nu)\lambda
\sigma_{k}\zeta(k,\nu)\beta(k,\nu)g_{\delta}(k,{\bf x})
\end{displaymath}

{\bf Proof}:if $z$ and $\zeta$ are real then  eq.(41) (cp. the
definition of ${\cal R}$ in eq.(40)) is a version of the Cameron-Martin formula
expressing the transformation of the Gaussian integral under a shift
of variables; the validity of the formula (41) for complex $z$ and
$\zeta$ follows by analycity, this has been discussed first in
ref.\cite{aze} and then in \cite{hab3}.

Now,we choose $z_{r}=(\lambda\sigma_{r})^{-1}$ and $\zeta(k,\nu)=
(\lambda\sigma_{k})^{-1}$.
Let us introduce a real field (see eq.(38) for a definition of ${\cal A}^{cl}$)
\begin{equation}
{\cal A}^{I}(\tau,{\bf x})\equiv{\cal A}^{cl}(\tau,{\bf x};
t,{\bf B},{\bf C}) +{\cal F}(\tau,{\bf x})
\end{equation}
and a real trajectory
\begin{displaymath}
{\bf x}(\tau)={\bf x}^{\prime}\frac{\tau}{t}+
\frac{t-\tau}{t}{\bf x}+ {\bf w}(\tau)
\end{displaymath}
Let us choose ${\bf w}_{r}$ in such a way that ${\bf x}_{r}(\tau) $ is the
 solution of the Lorentz equation
\begin{equation}
\begin{array}{l}
m_{r}\frac{d^{2}{\bf x}_{r}}{d\tau^{2}}=-e_{r}\nabla_{r}\sum_{s}
e_{s}\phi({\bf x}_{r},{\bf x}_{s})-\nabla_{r}u_{r}({\bf x}_{r})-
\cr
-\frac{1}{c}\partial_{\tau}{\cal A}^{I}(\tau,{\bf x}_{r})
+e_{r}\frac{d{\bf x}_{r}}{dt}
\times \nabla \times {\cal A}^{I}(\tau,{\bf x}_{r})
\end{array}
\end{equation}
with the two-point boundary condition ${\bf x}(0)={\bf x}$ and ${\bf x}(t)=
{\bf x}^{\prime}$. For a sufficiently small time t (what we assume here)
the two-point boundary value problem (43) has the unique solution (\cite{bai}).

The field ${\cal A}^{I}$ is constructed from the amplitudes
$\beta(\tau;k,\nu)$. Let us define
\begin{equation}
\xi(\tau;k,\nu)= \alpha(\tau;k,\nu) +\beta(\tau;k,\nu)
\end{equation}
where $\alpha$ is defined below eq.(34).Then, ${\cal A}^{I}$ of eq.(42)
is defined by $\xi$
\begin{displaymath}
{\cal A}^{I}(\tau,{\bf x})=\sum_{k,\nu}\sqrt{4\pi\mu_(k)}cf(k){\cal E}(k,\nu)
\xi(\tau;k,\nu)g(k,{\bf x})
\end{displaymath}
Now, we choose $\beta(\tau;k,\nu)$ in such a way that $\xi(\tau;k,\nu)$ is
 the solution of the equation
\begin{equation}
(\partial_{\tau}^{2}+\omega(k)^{2})\xi(\tau;k,\nu)=
\sqrt{\frac{4\pi}{\mu(k)}}f(k)\sum_{r}
\frac{d{\cal E}(k,\nu){\bf x}_{r}(\tau)}{d\tau}g(k,{\bf x}(\tau))
\end{equation}
with the two-point boundary condition $\xi(0;k,\nu)=X(k;\nu)$ and
$\xi(t;k,\nu)=Y(k;\nu)$.
Eq.(45) implies the usual equation for a vector potential of
a given particle current
\begin{equation}
(\frac{1}{c^{2}}\frac{\partial^{2}}{\partial\tau^{2}}-\triangle){\cal A}^{I}
({\bf x})
=\frac{4\pi}{c}
\sum_{r}e_{r}\frac{d{\bf x}_{r}}{d\tau}\delta_{f}({\bf x}_{r}(\tau),{\bf x})
\end{equation}
where
\begin{displaymath}
\delta_{f}({\bf x},{\bf y})=\frac{1}{V}\sum_{k}f(k)^{2}g(k,{\bf x})g(k,{\bf y})
\end{displaymath}
Eq.(46) is solved with the two-point boundary condition
 ${\cal A}^{I}(\tau=0,{\bf x})
={\bf B}({\bf x})$ and ${\cal A}^{I}(\tau=t,{\bf x})={\bf C}({\bf x}) $.
When $f(k)\rightarrow 1 $ then we obtain the classical electrodynamics
of point charges.We wrote eqs.(43) and (46) without the $\delta$-
regularization. As a first step we insert again the $\delta$-regularization
and prove the semiclassical expansion. Subsequently, we show that the
$\delta$-regularization can be removed.

{\bf Theorem 9}

Let us assume that V is a cube and the number of modes is finite.
Define the classical action (where ${\bf x}$ and $\xi$ solve the
classical equations of motion (43) and (45))
\begin{equation}
\begin{array}{l}
S_{\delta}=\sum_{r}\int_{0}^{t}\Big(\frac{1}{2m_{r}}(\frac{d{\bf x}_{r}}
{d\tau})^{2}
-u_{r}({\bf x}_{r}(\tau))\Big)d\tau -\sum_{r,s}\int_{0}^{t}e_{r}e_{s}
\phi_{\delta}({\bf x}_{r},{\bf x}_{s})d\tau+
\cr
+\sum_{k,\nu}\int_{0}^{t}
\big(\frac{1}{2\mu(k)}(\frac{d\xi(k,\nu)}{dt})^{2}
 -\frac{1}{2}\mu(k)\omega(k)^{2}\xi(k,\nu)^{2}\big)d\tau
 +\sum_{r}\frac{e_{r}}{c}
 \int_{0}^{t}{\cal A}_{\delta}^{I}(\tau,{\bf x}_{r}(\tau))
 d{\bf x}_{r}(\tau)
\end{array}
\end{equation}
then

i)for a sufficiently small time t
\begin{equation}
K_{\delta}(t,{\bf C},{\bf x};0,{\bf A},
{\bf x}^{\prime})exp(-i\frac{S_{\delta}}{\hbar})
\equiv\Omega_{\delta}(t,\hbar)
\end{equation}
is bounded in $\hbar$

ii) the limit $S_{\delta}\rightarrow S$ as $\delta \rightarrow 0$
exists if t is small enough

iii) for a sufficiently small time $\Omega_{\delta}$
 can be expanded in a power series in the electric charge
and the limit $\delta \rightarrow 0$ of each term exists uniformly in $\hbar$.
Hence, the limits $\delta \rightarrow 0$ and $\hbar \rightarrow 0$  in
$\Omega_{\delta}(t,\hbar) $ exist and
are exchangeable.

{\bf Proof}: We expand $K_{\delta}$ in powers of $\sqrt{\hbar}$
according to eqs.(40)-(41).$S_{\delta}$ of eq.(47) constitutes the singular
part of the expansion.Equations of motion (43) and (45) result from the
condition of the cancellation of the terms of order $(\sqrt{\hbar})^{-1}$.
There remains to bound the remainder. In the cube $g(k,{\bf x})$ are
the sinus and cosinus functions. In such a case the problem of the estimate
of the remainder is very similar to the estimate of the remainder for
trigonometric interactions done in 
ref.\cite{hab3}(Theorem 6.2).So, after a subtraction
of the term $S_{\delta}$ from $K_{\delta}$ we expand the exponentials in
eq.(41) in a  power series in the electric charge. The series is convergent
for a sufficiently small time.
We show that the singular terms of order $(\sqrt{\hbar})^{-1}$
cancel identically in the series if the classical equations of motion
(43) and (45) are fulfilled. The remaining terms in the series are
already bounded in $\hbar$ and $\delta$.As a consequence, the series
is convergent uniformly in $\hbar$ and $\delta$. Hence, the limit
$\delta \rightarrow 0$ can be performed term by term in the
convergent  power series. The part ii) of the theorem
is an easy consequence
of the continuity of solutions of differential equations
with respect to variation of parameters if the fields are regular
and the time is small enough. The part iii) follows from our discussion of
the expansion of $\Omega_{\delta}$ in a power series.

We can solve the linear equation (45) and insert the solution into
eq.(43).In such a case we obtain an integro-differential equation
describing particles interacting by Coulomb forces and additionally
under the interaction of the retarded
magnetic field produced by their movements. We would obtain this equation
if we performed the classical limit in eq.(27) where the
functional integration over the quantum electromagnetic field has
been calculated.The double stochastic integral in the exponential
of eq.(27) describes the quantum radiation reaction. Eqs. (43)
and (45) lead to the classical expression for the radiation reaction
 \cite{jac}\cite{roh}. We have shown
in Theorem 9 that the quantum radiation reaction has a classical
limit for a sufficiently small time. It is well known that
we encounter serious acausal effects with both the classical
and quantum radiation reaction (see a clarification of some
difficulties in ref.\cite{sha}). However, these problems
involving  a large time behaviour are beyond the scope of this
paper.
\section{Some approximations in QED}
It is quite difficult to obtain  non-perturbative results in QED
without further approximations. It is often argued  (see e.g.
\cite {dir},\cite{lui}) that because the electron in an atom is
localized around the nucleus and the wave length involved in the
light-atom interaction is large in comparison to the atom size
 then in quantum radiation theory
we may neglect the spatial dependence of quantum electromagnetic
field (such an assumption is also applied in mathematical physics papers
\cite{fro}\cite{spo}). 
Although this argument is convincing classically it would
be very difficult to justify it in a complete relativistic quantum
 theory of the electromagnetic field. In a relativistic theory
 the propagator D in eq.(27) is singular on the light cone.Hence,
 an expansion in coordinates of this propagator in eq.(27) could
 not be justified.We could consider the ultraviolet cutoff in
 D as an expression of the believe that the high
 frequency atom-photon interactions are not relevant to atomic
 physics.

We intend to formulate this short range approximation in
a particular case.Let us consider the formfactor
\begin{equation}
f(k)^{2}=\int_{0}^{\infty}ds \varsigma(s)exp(-\epsilon s|{\bf k}|)
\end{equation}
We wish to compare the transition amplitude (27) with the one
where the spatial dependence of D is neglected.
Let us note that (see the propagator D of Proposition 6)
\begin{equation}
\begin{array}{l}
D({\bf x},{\bf x}^{\prime};|\tau-\tau^{\prime}|)-
D(0,0;|\tau-\tau^{\prime}|)=
\cr
=\frac{1}{4\pi}\delta^{tr}
\int_{0}^{\infty}\varsigma(s)({\bf x}-{\bf x}^{\prime})^{2}
(\epsilon s + ic|\tau-\tau^{\prime}|)^{-2}
\left((\epsilon s + ic|\tau-\tau^{\prime}|)^{2}+ ({\bf x}-{\bf x}^{\prime}
)^{2}\right)^{-1}
\end{array}
\end{equation}
Hence, this difference can be a bounded function of ${\bf x}-{\bf x}^{\prime}$.

{\bf Proposition 10}

Consider the transition amplitude (27) for a single particle in $R^{3}$.
Assume that
 $\int_{0}^{\infty}s^{-2}\varsigma(s)ds <\infty$.Then, for a sufficiently small
time t
\begin{equation}
|(\Phi_{1}\chi,U_{t}\Phi_{2}\chi)- (\Phi_{1}\chi,U_{t}\Phi_{2}\chi)_{0}|
\leq const \frac{e^{2}}{\hbar c}t
\end{equation}
where $(,)_{0}$ denotes the transition amplitude computed with the propagator
$D(0,0;|\tau-\tau^{\prime}|)$ where the spatial dependence is neglected.

{\bf Proof}: We note that under the assumptions of the Proposition
the difference of the propagators $D(\lambda\sigma {\bf b}(\tau),
\lambda\sigma{\bf b}(\tau^{\prime});|\tau-\tau^{\prime}|)- D(0,0;|\tau-
\tau^{\prime}|)$ is a bounded function of ${\bf b}$ and $\tau-\tau^{\prime}$.
Then, in eq.(27) the exponential of the double stochastic integral
$\int dbdb(D-D_{0})$ (where the propagator
without the spatial dependence is denoted $D_{0}$) can be expanded in a
power series, which is convergent for a sufficiently small time.
The first term $\int dbdb(D-D_{0})$ in the series (51) is estimated by
the r.h.s. of eq.(51).
Then, the estimate of the proposition is a consequence of
the convergence of the series.

The proposition says that under our assumptions an error of the neglect of
the spatial dependence is of the order of the fine structure constant.
Such a result can be considered as a first step of a more systematic
investigation. The interesting result should hold true for arbitrarily
large time.However, this would be quite difficult to prove.The difficulty
results from the problem of integrability for arbitrarily large time
of the exponential of the double stochastic integral. The approximations
to the double stochastic integral are interesting from the point of view
of the quantum theory of radiation reaction.The well-known
formula for classical radiation reaction of a point charge \cite{jac}
confirmed by approximate quantum calculations  
\cite{sha}\cite{coh} can be obtained
by a neglect of spatial dependence. In order to investigate the problem
from the point of view of the Feynman integral (for a Hamiltonian
formulation  see \cite{ara} ) let us introduce the
decomposition $D=D^{0}+D^{1}$ for the zero temperature propagator and
$G=G^{0}+G^{1}$ for the finite temperature
propagator (22), where the index zero means that the spatial coordinates
are set equal to zero. In the infinite volume we have ( the factor
$\frac{2}{3}$ comes from the replacement of $\delta^{tr}$ by $\delta$)
\begin{displaymath}
D^{0}_{jl}(t)= \frac{2}{3}\delta_{jl}
\frac{\hbar c}{\pi}\int_{0}^{\infty}dk k f(k)^{2}
exp(-ickt)
\end{displaymath}
and
\begin{displaymath}
G^{0}(t)_{jl}=\frac{2}{3}\delta_{jl}
\frac{\hbar c}{\pi}\int_{0}^{\infty}dk k f(k)^{2}
\left(cth(\frac{1}{2}\beta\hbar c k)
cos(ckt) -isin(ckt)\right)
\end{displaymath}
When $f(k)\rightarrow 1$ then  $ImD^{0}(t)_{jl}\rightarrow \frac{2}{3c}\hbar
\delta_{jl}\delta^{\prime}(t)$ (where Im denotes the imaginary part).
Inserting this result into eq.(27) we obtain the standard expression for
the radiation reaction (strictly speaking the integral (27) becomes
divergent and requires a renormalization). We can take the radiation
reaction into account explicitly by means of a transformation of paths.
Let us introduce an operator in $L^{2}([0,t])$
\begin{displaymath}
{\cal K}=1 + \frac{e^{2}}{\hbar c^{2}m} ImD^{0}
\end{displaymath}
where by $ImD^{0}$ we denoted the integral operator defined by the
kernel $Im D^{0}$. A transformation
\begin{displaymath}
d{\bf b}\rightarrow d\underline{\bf b}\equiv{\cal K}^{\frac{1}{2}}d{\bf b}
\end{displaymath}
expresses the functional integral (27) in terms of a new Brownian
motion which contains the effect of the radiation reaction.
Such a transformation is mathematically correct if $Im D^{0}$ is a
Hilbert-Schmidt operator and moreover if ${\cal K}$ is positive
definite.For positive temperature it is useful to note that
$G-D$ is real.So, the heat bath does not contribute to the classical
radiation reaction. In order to see the effect of the temperature we
perform another transformation upon the functional integral.
For any  functional $\Phi$ of the Brownian motion defined on the
interval [0,t] we have the Cameron-Martin identity
\begin{displaymath}
E[\Phi({\bf b})]=E\bigg[exp\Big(-\int_{0}^{t}\frac{d{\bf f}}
{d\tau}d{\bf b}(\tau)
-\frac{1}{2}\int_{0}^{t}(\frac{d{\bf f}}{d\tau})^{2}d\tau\Big)
\Phi({\bf b}+{\bf f})\bigg]
\end{displaymath}
for any ${\bf f}$ with a square integrable derivative.We choose now
\begin{displaymath}
\frac{d{\bf f}}{d\tau}=\frac{ie}{\hbar c}\lambda\sigma{\cal A}^{z}(0,\tau)
\end{displaymath}
in eq.(24)( for simplicity restricted to a single particle).In such a
case in eq.(24) $\Phi_{2}({\bf x}+\lambda\sigma{\bf b}(t))\rightarrow
\Phi_{2}({\bf q}_{t}({\bf x}))$ where  ${\bf q}_{t}({\bf x})$ is the
solution of the equation
\begin{equation}
d{\bf q}=-\frac{e}{m} {\cal A}^{z}(0,\tau)d\tau +\lambda\sigma d{\bf
b}(\tau)
\end{equation}
Using eqs.(52) and (22) we can compute the correlation functions
of q. In this way for a large temperature and $f(k)=1$ we obtain
\begin{displaymath}
E[\frac{dq_{j}}{d\tau}\frac{dq_{l}}{d\tau^{\prime}}]=\frac{e^{2}}{m^{2}}
KT\delta_{jl}\delta(\tau-\tau^{\prime})
\end{displaymath}
plus less singular terms. So, the charged particle in equilibrium
with a black body radiation behaves like a Brownian particle in a
fluid of temperature T. This result is approximate, because
we neglected the change of the covariance of the electromagnetic
field caused by the Cameron-Martin transformation. We could
approach this way in the framework of QED the model discussed
first by Einstein and Hopf \cite{ein} and later on in refs.\cite{boy1}
\cite{mil}.Let us note that the behaviour of the correlation functions
of the quantum particle in equilibrium with the radiation is
in qualitative agreement with the requirements of the "quantum
Brownian motion " due to Benguria and Kac \cite{ben}( see also
ref.\cite{leg2}).

\section{Discussion and outlook}

In this paper we concentrated mainly on the precise
formulation of regularized QED dynamics.
In some  cases we were able to show that the expressions
were finite only for a small time. Such a requirement resulted
from a demand of a convergence of a perturbation series.
In principle, we can prolong the dynamics to a larger time
using the group property of the time evolution. In practice,
the multiple integration involved in such a procedure may be
inconvenient.
The results of sec.7 may indicate a way out of this difficulty.
The formula for the propagation kernel of a harmonic oscillator
holds true for arbitrarily large time (except of the moments equal
to the classical periods) in spite of the fact that the
functional integral looses its sense already when t tends to the
first period. In another approach to the harmonic
 oscillator we have obtained
in sec.3 the dynamics for an arbitrarily large time through an application
of stochastic differential equations (see eq.(5)). We have
discussed the method of stochastic differential equations for
non-linear quantum systems in refs.\cite{hab2}\cite{hab4}. This
method could be applied to QED. The first step has been done in eq.(52).
A non-linear transformation (discussed for quantum mechanics in
\cite{hab4}) can be applied to formulate exactly the dynamics in
QED as a random dynamics. We intend to discuss this method in a
subsequent publication.

The dependence on the formfactor (ultraviolet
cutoff) constitutes another unsatisfactory aspect of the theory.
 A non-perturbative  removal of the cutoff
 is a difficult problem. However, some results may be insensitive
 to the cutoff. One may believe that phenomena on a large time and
 space scale involving low energies should not depend on the
 ultraviolet cutoff. We consider the formulation of this paper
 as a substantial step towards a control over approximations
 made in QED, quantum optics and quantum radiation theory.
 In principle, in sec.9 we were able to estimate an
 error involved in a dipole approximation. In sec.8 we are able
 to estimate subsequent terms in the semiclassical expansion.

\end{document}